\documentclass[11pt,a4paper]{article}

\usepackage[margin=2cm]{geometry}
\usepackage{amssymb,amsfonts,amsmath}
\numberwithin{equation}{section}
\usepackage{cite,enumerate,float}
\usepackage{color}
\usepackage{tikz}
\usetikzlibrary{arrows,snakes,backgrounds}
\usepackage[colorlinks=true,
linkcolor=blue,
citecolor=blue,
urlcolor=blue]{hyperref}
\usepackage{authblk}

\def\be{\begin{eqnarray}}
\def\ee{\end{eqnarray}}
\def\nn{\nonumber}

\def\p{\partial}

\def\Tr{{\rm Tr}\,}

\newcommand{\ri}{\mathrm{i}}

\newcommand{\hf}{\frac{1}{2}}

\newcommand{\bra}{\langle}
\newcommand{\ket}{\rangle}
\newcommand{\la}{\lambda}

\newcommand{\al}{\alpha}

\newcommand{\rt}[1]{\sqrt{#1}}

\newcommand{\cZ}{\mathcal{Z}}

\newcommand{\cA}{\mathcal{A}}

\newcommand{\cN}{\mathcal{N}}

\definecolor{red}{rgb}{1,0,0}
\definecolor{orange}{rgb}{1,0.5,0}
\definecolor{violet}{rgb}{0.7,0,1}

\begin{document}

\setlength{\baselineskip}{6mm}

\title{\LARGE \textbf{Group character averages via a single Laguerre}}
\author[1,2,3,4]{Alexei Morozov}
\author[5]{Kazumi Okuyama}

\affil[1]{MIPT, Dolgoprudny, 141701, Russia}
\affil[2]{NRC Kurchatov Institute 123182, Moscow, Russia}
\affil[3]{Institute for Information Transmission Problems, Moscow 127994, Russia}
\affil[4]{ITEP, 117259, Moscow, Russia}

\affil[5]{Department of Physics,
Shinshu University,\authorcr
3-1-1 Asahi, Matsumoto 390-8621, Japan}

\date{}

\bigskip

\maketitle

\vspace{-9cm}

\begin{center}
  \hfill MIPT/TH-08/26\\
  \hfill ITEP/TH-08/26\\
  \hfill IITP/TH-08/26
\end{center}

\vspace{6.5cm}

\begin{abstract}
 Average of exponential ${\rm Tr}_R e^X$, i.e. of a group rather than an algebra character,
in Gaussian matrix model is known to be an amusing generalization of Schur polynomial,
where time variables are substituted by traces of products of non-commuting matrices
${\rm Tr} \left(\prod_i A_{k_i}\right)$  and are thus labeled by weak compositions.
The entries of matrices $A_k$ are made from extended Laguerre polynomials,
what introduces additional difficulties.
We describe the generic sum rules, which  express arbitrary traces
through convolutions of a single Laguerre polynomial $L_{N-1}^1(z_{k_i})$,
what is a considerable simplification.
\end{abstract}


%
%
%
%
%

\bigskip

\bigskip

\section{Introduction}
Random matrix model is a ubiquitous tool in physics and mathematics.
Of particular importance is the Gaussian matrix model
\begin{equation}
\begin{aligned}
 \cZ=\int dX e^{-\hf \Tr X^2},
\end{aligned}
\label{eq:gaussian}
\end{equation}
where $X$ is an $N\times N$ hermitian random matrix.
This matrix model plays an important role in the study of the moduli space of
Riemann surfaces \cite{Zagier1986,norbury2008counting}.
In particular, one can compute the Euler characteristic and the discrete
volume of the moduli space of
Riemann surfaces from the correlators of $\Tr X^k$'s.
\begin{equation}
\begin{aligned}
\left\bra\prod_{i=1}^m\Tr X^{k_i}\right\ket=\frac{1}{\cZ}
\int dX e^{-\hf\Tr X^2}\prod_{i=1}^m\Tr X^{k_i}.
\end{aligned}
\label{eq:corr}
\end{equation}
As discussed in \cite{Morozov:2009uy}, we can study the
connected part of the correlator \eqref{eq:corr} by
introducing various generating functions of $\Tr X^k$.
In this paper, we consider the exponential $\Tr e^{sX}$ of $X$
and their
connected $m$-point correlator
\begin{equation}
\begin{aligned}
 \left\bra\prod_{i=1}^m\Tr e^{s_iX}\right\ket_c,
\end{aligned}
\label{eq:exp-corr}
\end{equation}
where the subscript $c$ refers to the connected part.
Interestingly, this type of correlator also appears in the
computation of the expectation value of supersymmetric Wilson loop
in $\cN=4$ super Yang-Mills
\cite{Erickson:2000af,Drukker:2000rr}
thanks to the supersymmetric localization
\cite{Pestun:2007rz}.\footnote{See also \cite{Muck:2019hnz,Muck:2021vyx}.}
Also, the exponential correlator \eqref{eq:exp-corr}
is important to understand the ramp and the plateau behavior
of the spectral form factor \cite{brezin-hikami,Okuyama:2018yep}.

The exponential correlator \eqref{eq:exp-corr}
has been studied in \cite{Morozov:2009uy,Morozov:2026jyc}
by utilizing the underlying (super)integrability of the matrix model.
However, this innocent-looking correlator
\eqref{eq:exp-corr} turns out to be extremely difficult to evaluate in a
closed form.
In this paper, we will take a modest step toward
the exact computation of \eqref{eq:exp-corr}.
As discussed in \cite{Okuyama:2018aij,Okuyama:2018yep},
the exponential correlator \eqref{eq:exp-corr}
is written as some combination of the trace of the $N\times N$ matrix $A(s)$
\begin{equation}
\begin{aligned}
 A(s)_{ij}=\bra i|e^{s(a+a^\dagger)}|j\ket=\rt{\frac{i!}{j!}}e^{\hf s^2}
s^{j-i}L_i^{j-i}(-s^2),\quad(i,j=0,\cdots, N-1),
\end{aligned}
\label{eq:Amat-with}
\end{equation}
where $a,a^\dagger$ is the harmonic oscillator $[a,a^\dagger]=1$
and $L_n^\al(x)$ denotes the Laguerre polynomial.
The harmonic oscillator naturally arises in this computation since
the orthogonal polynomial with respect to the Gaussian
measure \eqref{eq:gaussian} is the Hermite polynomial, which of course is the wavefunction of the harmonic oscillator.
For instance, one- and two-point connected correlator of $\Tr e^{sX}$ are given by
\begin{equation}
\begin{aligned}
 \bra \Tr e^{sX}\ket&=\Tr A(s),\\
\bra \Tr e^{s_1X}\Tr e^{s_2X}\ket_c&=\Tr \bigl[A(s_1+s_2)-A(s_1)A(s_2)\bigr].
\end{aligned}
\label{eq:1-2-corr}
\end{equation}
The one-point function $\Tr A(s)$ has been evaluated in a closed form in
\cite{Drukker:2000rr}
\begin{equation}
\begin{aligned}
 \Tr A(s)=e^{\hf s^2}L_{N-1}^1(-s^2).
\label{eq:DG}
\end{aligned}
\end{equation}
We find that $\Tr A(s_1)\cdots A(s_m)$ is written as a
convolution of $L_{N-1}^1$'s
\begin{equation}
\begin{aligned}
 \Tr A(s_1)\cdots A(s_m)&=
\frac{1}{\sum_{k=1}^m s_k}
\prod_{j=1}^{m-1}\int_0^\infty dx_j e^{-s_jx_j}
\prod_{k=1}^m s_ke^{\hf s_k^2}L_{N-1}^1\bigl(-s_k^2+s_kx_{k}-s_kx_{k-1}\bigr)\\
&\qquad+\text{cyclic permutations of}~(s_1\cdots s_m),
\end{aligned}
\label{eq:tr-conv}
\end{equation}
with $x_0=x_m=0$, and we assumed $s_i>0$ in \eqref{eq:tr-conv}. In the second line,
we cyclically permute the parameters $(s_1\cdots s_m)$ only; we do not permute the integration
variables $(x_1\cdots x_{m-1})$.
As discussed in \cite{Okuyama:2018aij},
this trace \eqref{eq:tr-conv}
is the basic building block of the
exponential correlator \eqref{eq:exp-corr}.
It is interesting to note that the
exponential correlator of the Airy matrix model
has a similar structure of the convolution \cite{Okounkov2001GeneratingFF};
this is not so surprising since the Airy matrix model
is obtained by a double scaling limit of the Gaussian matrix model
(see e.g. \cite{Ginsparg:1993is} for a review).

This paper is organized as follows. 
In section \ref{sec:A1}, we compute the trace
\eqref{eq:tr-conv} for the case where all parameters $s_i$ are equal. 
In section \ref{TrueTr}, we compute the trace \eqref{eq:tr-conv}
with generic parameters $s_i$. 
In section \ref{sec:conn}, we consider the connected correlator of exponentials
$\Tr e^{sX}$ from the viewpoint of superintegrability in \cite{Morozov:2026jyc}.
In section \ref{sec:comparison}, we compare our approach in the present paper
with \cite{Morozov:2009uy}.
Finally, we conclude in section \ref{sec:conclusion}
with some discussion on the future problems.

\section{$\Tr (\cA_1)^m$ via Laguerres}\label{sec:A1}
In this section, we consider the trace
\eqref{eq:tr-conv} for the case where all parameters $s_i$ are equal.
For simplicity, it is sometimes convenient to omit the 
exponential factor $e^{\hf s^2}$ in \eqref{eq:Amat-with}
and use the matrix $\mathcal{A}(s)\equiv e^{-\hf s^2}A(s)$.
More explicitly, the matrix element of $\mathcal{A}(s)$ is given by
\begin{equation}
\begin{aligned}
 \mathcal{A}(s)_{ij}=
\rt{\frac{i!}{j!}}
s^{j-i}L_i^{j-i}(-s^2).
\end{aligned} 
\label{eq:cA-def}
\end{equation}
Note that $\mathcal{A}(s)$ is a symmetric matrix: 
$\mathcal{A}(s)_{ij}=\mathcal{A}(s)_{ji}$.
We also use the notation $\mathcal{A}_k\equiv\mathcal{A}(ks)$.
In this section, we consider $\Tr (\cA_1)^m$ as a warm-up
for the computation of \eqref{eq:tr-conv}.
From the structure of $\cA$ in \eqref{eq:cA-def},
one can easily see that the factor of $\rt{i!/j!}s^{j-i}$ drops out in
the trace $\Tr (\cA_1)^m$ 
\begin{equation}
\begin{aligned}
 \Tr ({\cal A}_1)^m = 
\sum_{i_1,\cdots,i_m=0}^{N-1} \prod_{k=1}^m L^{i_{k+1}-i_k}_{i_k}(z),
\end{aligned} 
\end{equation}
where $i_{m+1} = i_1$ and $z=-s^2$.
In this sense $\Tr (\cA_1)^m$ is made from Laguerre polynomials from the very beginning,
but seems to involve polynomials $L^\al_n(z)$ of different levels $\al$, from $-N+1$ to $N-1$.
It turns out that this trace can be rewritten through those with $\al=1$ only.

Making use of the generating function for Laguerre polynomials,
\be
\frac{e^{-\frac{tz}{1-t}}}{(1-t)^{\al+1}} = \sum_{n=0}^\infty L^\al_n(z) t^n,
\label{eq:gen-L}
\ee
we obtain:
\paragraph{${\bf m=1}$:}
\be
\Tr {\cal A}_1 =
\sum_{i=0}^{N-1}  L^0_{i}(z)
= \oint\frac{dt}{2\pi\ri}  \frac{e^{-\frac{tz}{1-t}}}{(1-t) } \sum_{i=0}^{N-1} 
\frac{1}{t^{i+1}}
= \oint\frac{dt}{2\pi\ri}  \frac{e^{-\frac{tz}{1-t}}}{(1-t)^2 } \left(\frac{1}{t^N}-\underline{1}\right)
= L^1_{N-1}(z).
\ee
Note that the underlined $\underline 1$ can be omitted,
since it gives rise to a series with non-negative powers of $t$ only,
which is annihilated by $\oint dt$.
\paragraph{${\bf m=2}$:}
\begin{equation}
\begin{aligned}
 \Tr ({\cal A}_1)^2 &=
\sum_{i,j=0}^{N-1}  L^{j-i}_{i}(z)L^{i-j}_{j}(z)= \oint\frac{dt}{2\pi\ri}\oint
\frac{dt'}{2\pi\ri}  \sum_{i,j =0}^{N-1}
\frac{e^{-\frac{tz}{1-t}- \frac{t'z}{1-t'}}}{(1-t)^{j-i+1}(1-t')^{i-j+1} }
 \frac{1}{t^{i+1}{t'}^{j+1}}\\
 &= \oint\frac{dt}{2\pi\ri}\oint\frac{dt'}{2\pi\ri} \frac{e^{-\frac{tz}{1-t}- \frac{t'z}{1-t'}}}{(1-t) (1-t')}\cdot
\frac{ \left(\frac{1-t}{t(1-t')}\right)^{N}-\underline{1}}{\frac{1-t}{t(1-t')}-1} \cdot
\frac{ \left(\frac{1-t'}{t'(1-t)}\right)^{N}-\underline{1}}{\frac{1-t'}{t'(1-t)}-1}
\\
&=  \oint\frac{dt}{2\pi\ri}\oint\frac{dt'}{2\pi\ri} \frac{e^{-\frac{tz}{1-t}- \frac{t'z}{1-t'}}}
{(1-2t-tt')(1-2t'+tt')}
\cdot \frac{1}{t^N{t'}^N},
\end{aligned} 
\end{equation}
%
where again the underlined terms do not contribute and can be omitted for the same reason.
Remarkably this is equal to a simple convolution of two Laguerre polynomials:
\begin{equation}
\begin{aligned}
 &\int_0^\infty dx e^{-x} L^1_{N-1}(z+x)L^1_{N-1}(z-x) \\
=&
\int_{0}^\infty dx e^{-x}\oint\frac{dt}{2\pi\ri}\oint\frac{dt'}{2\pi\ri} \frac{e^{-\frac{t(z+x)}{1-t}- \frac{t'(z-x)}{1-t'}}}{(1-t)^2(1-t')^2}
\cdot \frac{1}{t^N{t'}^N}\\
=& \oint\frac{dt}{2\pi\ri} \oint\frac{dt'}{2\pi\ri} \frac{e^{-\frac{tz}{1-t}- \frac{t'z}{1-t'}}}{(1-t)(1-t')(1-2t'+tt')}
\cdot \frac{1}{t^N{t'}^N}\\
=& \frac{1}{2} \oint\frac{dt}{2\pi\ri} \oint
\frac{dt'}{2\pi\ri}  \frac{e^{-\frac{tz}{1-t}-\frac{t'z}{1-t'}}}{(1-t)(1-t')}
\left(\frac{1}{1-2t'+tt'} + \frac{1}{1-2t+tt'}\right)
\cdot \frac{1}{t^N{t'}^N}\\
=&  \oint\frac{dt}{2\pi\ri} \oint\frac{dt'}{2\pi\ri}  \frac{e^{-\frac{tz}{1-t}- \frac{t'z}{1-t'}}}
{(1-2t+tt')(1-2t'+tt')}
\cdot \frac{1}{t^N{t'}^N},
\end{aligned} 
\end{equation}
where we made use of the obvious symmetry between $t$ and $t'$.
Thus
\be
\boxed{
\Tr ({\cal A}_1)^2 =
\int_0^\infty dx e^{-x} L^1_{N-1}(z+x)L^1_{N-1}(z-x)
}
\label{eq:A1^2}
\ee
is indeed expressed through $L^1_{N-1}$ only.
\paragraph{${\bf m=3}$:}
\begin{equation}
\begin{aligned}
 \Tr ({\cal A}_1)^3 &=
\sum_{i,j,k=0}^{N-1}  L^{j-i}_{i}(z)L^{k-j}_{j}(z)L^{i-k}_{k}(z)\\
&= \oint\frac{dt}{2\pi\ri}\oint\frac{dt'}{2\pi\ri}\oint\frac{dt''}{2\pi\ri}  \sum_{i,j,k =0}^{N-1}
\frac{e^{-\frac{tz}{1-t}- \frac{t'z}{1-t'}- \frac{t''z}{1-t''}}}{(1-t)^{j-i+1}(1-t')^{k-j+1}(1-t'')^{i-k+1} }
 \frac{1}{t^{i+1}{t'}^{j+1}{t''}^{k+1}}\\
& = \oint\frac{dt}{2\pi\ri}\oint\frac{dt'}{2\pi\ri}\oint\frac{dt''}{2\pi\ri}
 \frac{e^{-\frac{tz}{1-t}- \frac{t'z}{1-t'}-\frac{t''z}{1-t''}}}{(1-t) (1-t')(1-t'')}\cdot
\frac{ \left(\frac{1-t}{t(1-t'')}\right)^{N}-\underline{1}}{\frac{1-t}{t(1-t'')}-1} \cdot
\frac{ \left(\frac{1-t'}{t'(1-t)}\right)^{N}-\underline{1}}{\frac{1-t'}{t'(1-t)}-1} \cdot
\frac{ \left(\frac{1-t''}{t''(1-t')}\right)^{N}-\underline{1}}{\frac{1-t''}{t''(1-t')}-1}\cdot \frac{1}{t{t'}{t''}}\\
&=  \oint\frac{dt}{2\pi\ri}\oint\frac{dt'}{2\pi\ri}\oint\frac{dt''}{2\pi\ri} \frac{e^{-\frac{tz}{1-t}- \frac{t'z}{1-t'}- \frac{t''z}{1-t''}}}
{(1-2t+tt'')(1-2t'+tt')(1-2t''+t't'')}
\cdot \frac{1}{t^N{t'}^N{t''}^N}.
\end{aligned} 
\end{equation}
Now we can convert
\begin{equation}
\begin{aligned}
 &\frac{1}{(1-2t+tt'')(1-2t'+tt')(1-2t''+t't'')}
= \frac{1}{3}\Biggl[
\frac{1}{(1-t)(1-t'')(1-2t'+tt')(1-2t''+t't'')} + \\
&\quad+ \frac{1}{(1-t)(1-t')(1-2t+tt'') (1-2t''+t't'')}
+ \frac{1}{(1-t')(1-t'')(1-2t+tt'')(1-2t'+tt') }
\Biggr].
\end{aligned} 
\label{tripleexpan}
\end{equation}
On the other hand, the trilinear combination of Laguerres, which we are looking for is
\begin{equation}
\begin{aligned}
 &L_{N-1}^1(z+y_1)L_{N-1}^1(z+y_2)L_{N-1}^1(z+y_3)\\
 =&\oint\frac{dt}{2\pi\ri}\oint\frac{dt'}{2\pi\ri}\oint\frac{dt''}{2\pi\ri}
e^{-\frac{ty_1}{1-t}- \frac{t'y_2}{1-t'}- \frac{t''y_3}{1-t''}}\cdot
\frac{e^{-\frac{tz}{1-t}- \frac{t'z}{1-t'}- \frac{t''z}{1-t''}}}{(1-t)^2(1-t')^2(1-t'')^2 }
\cdot \frac{1}{t^N{t'}^N{t''}^N}.
\end{aligned} 
\label{tripleL}
\end{equation}
Now we need to get the triple brackets in (\ref{tripleexpan}) from $x$-integrations.
For example, the first term can be obtained as follows:
\begin{equation}
\begin{aligned}
 &\frac{1}{(1-t)(1-t'')(1-2t'+tt')(1-2t''+t't'')} \\
&= \frac{1}{(1-t)^2(1-t')^2(1-t'')^2} \int_0^\infty  dx_1
e^{-x_1\left(1+\frac{t}{1-t} - \frac{t'}{1-t'}\right)}
\int_0^\infty dx_2  e^{-x_2\left(1+\frac{t'}{1-t'} - \frac{t''}{1-t''}\right)},
\end{aligned} 
\end{equation}
which means that in (\ref{tripleL}) we should put $y_1= x_1$, $y_2 = x_2-x_1$, 
$y_3=-x_2$.
Due to cyclic symmetry, 
the other two terms are represented exactly in the same way.
This means that
\be
\boxed{\Tr ({\cal A}_1)^3 
=   \int_0^\infty \int _0^\infty dx_1dx_2 e^{-x_1-x_2}
L^1_{N-1}(z+x_1)L^1_{N-1}(z+x_2-x_1)L^1_{N-1}(z-x_2).}
\ee
Again we obtained an expression in terms of a single Laguerre $L^1_{N-1}$.
Generalizations are straightforward.
For instance, for 
$m=4$ we find
\begin{equation}
\begin{aligned}
 \Tr ({\cal A}_1)^4
=   \int_0^\infty\prod_{i=1}^3 
dx_i e^{-x_i}
L^1_{N-1}(z+x_1)L^1_{N-1}(z+x_2-x_1)L^1_{N-1}(z+x_3-x_2)L^1_{N-1}(z-x_3),
\end{aligned} 
\end{equation}
and so on.

\section{Generic traces of the matrices $\cA(s)$}\label{TrueTr}
In the previous section, we considered the trace $\Tr(\cA_1)^m=\Tr \cA(s)^m$.
Now we switch to traces of the  
product of matrices $\cA(s)$ with generic parameters $(s_1\cdots s_m)$.

Let us consider $\Tr  \cA(s_1)\cA(s_2)$
as an example. By a similar computation of the previous section,
we find that \eqref{eq:A1^2} gets substitute by a slightly more complicated
expression
\begin{equation}
\begin{aligned}
 \Tr  \cA(s_1)\cA(s_2)
=& \sum_{i,j =0}^{N-1}  s_1^{j-i}L^{j-i}_{i}(-s_1^2)
s_2^{i-j}L^{i-j}_{j}(-s_2^2) \\
=& \frac{s_1s_2}{s_1+s_2}
\int _0^\infty dx \Bigl[e^{-xs_1}L^1_{N-1}(-s_1^2+xs_1)L^1_{N-1}(-s_2^2-xs_2)+
(s_1\leftrightarrow s_2)\Bigr].
\end{aligned} 
\label{trAA }
\end{equation}
For $s_1=s_2=s$ we return to  \eqref{eq:A1^2}, 
with the obvious change $z=-s^2$.

Generalizations are straightforward.
For instance, the trace of three $\cA(s)$'s is given by
\begin{equation}
\begin{aligned}
 \Tr \cA(s_1)\cA(s_2)\cA(s_3)
=&\frac{s_1s_2s_3}{s_1+s_2+s_3}
\int_0^\infty dx_1\int _0^\infty  dx_2\\
\times&
\Bigl[G(s_1,s_2,s_3|x_1,x_2)+G(s_2,s_3,s_1|x_1,x_2)+G(s_3,s_1,s_2|x_1,x_2)\Bigr],
\end{aligned} 
\label{eq:trAAA}
\end{equation}
with
\be
G(s_1,s_2,s_3|x_1,x_2)=
e^{-x_1s_1-x_2s_2}
L^1_{N-1}(-s_1^2+x_1s_1)L^1_{N-1}(-s_2^2+x_2s_2-x_1s_2)L^1_{N-1}(-s_3^2-x_2s_3).
\ee
Note that only cyclic permutations are included
in \eqref{eq:trAAA}.
Note also that cyclically permuted are ``external'' $s$-variables, while integration $x$-variables are not affected.

Because of symmetricity of the matrix \eqref{eq:cA-def} 
the trace of a product of three $\cA$'s
is fully symmetric, not only cyclically
\begin{equation}
\begin{aligned}
 \Tr \cA(s_1)\cA(s_2)\cA(s_3)&=
\Tr \bigl[\cA(s_1)\cA(s_2)\cA(s_3)\bigr]^T\\
&=\Tr \cA(s_3)^T\cA(s_2)^T\cA(s_1)^T
=\Tr \cA(s_3)\cA(s_2)\cA(s_1).
\end{aligned} 
\label{eq:sym-AAA}
\end{equation}
Thus the non-commutativity of $\cA(s)$ with different $s$ does not affect 
the traces of three $\cA$'s.\footnote{This explains the ``puzzle'' in \cite{Morozov:2026jyc}, where non-symmetric version of ${\cal A}$ was used.}
This non-commutativity starts affecting traces of  higher powers, 
beginning from $m=4$.

Generalization of \eqref{eq:trAAA} to higher powers of $\cA(s_k)$'s is straightforward.
For the trace with generic parameters $(s_1\cdots s_m)$, 
we find the result \eqref{eq:tr-conv}.\footnote{One can prove \eqref{eq:tr-conv} by using the following identities:
\begin{equation}
\begin{aligned}
 \prod_{k=1}^m\frac{1}{1-t_{k+1}-\frac{s_{k+1}}{s_k}t_{k+1}(1-t_k)}
&=\frac{1}{s_1+\cdots +s_m}
\sum_{l=1}^m\frac{s_l}{(1-t_l)(1-t_{l+1})}
\prod_{\substack{k=1\\k\ne l}}^m
\frac{1}{1-t_{k+1}-\frac{s_{k+1}}{s_k}t_{k+1}(1-t_k)},\\
\int_0^\infty dx_k e^{-\bigl(s_k+\frac{t_k}{1-t_k}s_k-
\frac{t_{k+1}}{1-t_{k+1}}s_{k+1}\bigr)x_k}&=
\frac{(1-t_k)(1-t_{k+1})}{s_k(1-t_{k+1})-s_{k+1}t_{k+1}(1-t_k)},
\end{aligned} 
\label{eq:ptep-id}
\end{equation}
where $t_{m+1}=t_1$ and $s_{m+1}=s_1$.
We would like to thank the anonymous referee of PTEP for suggesting 
the use of \eqref{eq:ptep-id} and \eqref{eq:ptep-id2}.
}
Note that, for the trace of $\cA(s_k)$'s, the factor of $e^{\hf s_k^2}$
should be removed from \eqref{eq:tr-conv}.
See also appendix \ref{app:alt} for an alternative derivation of \eqref{eq:tr-conv}.

%
%
%

\section{Connected correlators}\label{sec:conn}

In the remaining two sections we consider combinations of different correlators.
Their normalization will matter, and we specify it once again:
we use the matrix $A(s)$ in \eqref{eq:Amat-with} instead of
$\cA(s)$ in \eqref{eq:cA-def}. We will also use the notation $A_k=A(ks)$
in what follows.

Correlators of various ${\rm Tr}_R e^{\cal X} =S_R\{p_k=\Tr e^{kX}\}$ were expressed through
traces of $A$ in \cite{Morozov:2026jyc}.
However, these expressions contain products of traces, what reflects the complicated
dependence of original correlators on fundamental traces.
Moreover, correlators of multiple traces, i.e. of products of Schur polynomials,
are expressed through Hall-Littlewood coefficients, what introduces additional complications.
At the same time, {\it connected} correlators of time-variables $\pi_m := \Tr e^{mX}$
should be represented as single traces -- and they actually are, for example,
from  (22)-(28) and  (41)-(44) of \cite{Morozov:2026jyc}\footnote{$\la$ in \cite{Morozov:2026jyc} corresponds to $s$ in this paper.}:
\begin{equation}
\begin{aligned}
 \langle\pi_1 \pi_1\rangle_c&:=
\langle\pi_1 \pi_1\rangle-\langle\pi_1\rangle^2 = \sigma_{[2]}+\sigma_{[1,1]}   - \sigma_{[1]}^2
= P_{[2]} e^{2s^2} + 2P_{[1,1]}e^{s^2}       -   P_{[1]}^2e^{s^2}\\
&= \Tr A_2 + (\Tr A_1)^2 -  \Tr (A_1^2)        -  (\Tr A_1)^2
 =   \Tr \Big(A_2  -  A_1^2\Big),\\
\langle\pi_2\pi_1\rangle_c&:=
\langle\pi_2\pi_1\rangle-\langle\pi_2\rangle\langle\pi_1\rangle
= \sigma_{[3]}-\sigma_{[1,1,1]}-\Big(\sigma_{[2]}-\sigma_{[1,1]}\Big) \sigma_{[1]}
\approx P_{[3]}+P_{[1,2]}  - P_{[2]} P_{[1]} \approx \\
&= \Tr A_3 + \Tr A_2\, \Tr A_1 - \Tr (A_2A_1) - \Tr A_2\, \Tr A_1
=  \Tr \Big(A_3 -  A_2A_1\Big),\\
\langle\pi_3\pi_1\rangle_c&:= \langle\pi_3\pi_1\rangle-\langle\pi_3\rangle\langle\pi_1\rangle
= \sigma_{[4]} - \sigma_{[2,2]} + \sigma_{[1,1,1,1]}
-\Big(\sigma_{[3]}-\sigma_{[1,2]}+\sigma_{[1,1,1]}\Big) \sigma_{[1]}
\approx P_{[4]}+P_{[1,3]}    -    P_{[3]} P_{[1]} \approx  \\
&=\Tr A_4 + \Tr A_3\, \Tr A_1 - \Tr (A_3A_1)  - \Tr A_3\, \Tr A_1
= \Tr \Big(A_4 -  A_3A_1\Big),\\
\ldots
\end{aligned} 
\end{equation}
(we omit exponentials of $s^2$ in the two last examples, what is denoted by $\approx$).
The final formula is expressed through $A$ from \eqref{eq:Amat-with}
with the exponential factor included.
We see that the products of traces drop out of these formulas,
In general for pair correlators
\be
\langle \pi_{s_1} \pi_{s_2} \rangle_c :=\langle \pi_{s_1} \pi_{s_2} \rangle
- \langle \pi_{s_1}\rangle \langle \pi_{s_2} \rangle =
\Tr\Bigl[A(s_1+s_2)-A(s_1)A(s_2)\Bigr],
\label{twopointconn}
\ee
which agrees with \eqref{eq:1-2-corr}.
Normalizations can be easily checked at $N=1$, where we obviously expect and obtain
$e^{\hf (s_1+s_2)^2} - e^{\hf s_1^2+\hf s_2^2}$.

Multipoint connected correlators can be defined through a logarithm of the generating functions:
\be
Z\{t\} = \langle e^{\sum_m t_m \pi^m}\rangle =1+
\sum_{n=1}^\infty \sum_{m_1,\cdots, m_n=1}^\infty
C_{m_1,\ldots,m_n} t_{m_1}\ldots t_{m_n} \langle \pi_{m_1}\ldots \pi_{m_n}\rangle,
\nn \\
\log Z\{t\} = \sum_{n=1}^\infty \sum_{m_1,\cdots, m_n=1}^\infty
c_{m_1,\ldots,m_n}t_{m_1}\ldots t_{m_n} \langle \pi_{m_1}\ldots \pi_{m_n}\rangle_c.
\ee
For example,
\begin{equation}
\begin{aligned}
 \langle \pi_1^3\rangle_c &:=
\langle \pi_1^3\rangle - 3\langle\pi_1^2\rangle \langle\pi_1\rangle  + 2\langle \pi_1\rangle^3
= \big(\sigma_{[3]}+2\sigma_{[1,2]}+\sigma_{[1,1,1]}\big)
- 3\big(\sigma_{[2]}+\sigma_{[1,1]}\big)\sigma_{[1]} + 2\sigma_{[1]}^3 \\
&\approx \Big(P_{[3]}+3P_{[1,2]} + 6 P_{[1,1,1]}\Big) - 3\Big(P_{[2]}+2P_{[1,1]}\Big)P_{[1]}+2P_{[1]}^3\\
&= P_{[3]}+3P_{[1,2]}+6P_{[1,1,1]}-3P_{[2]}P_{[1]} -6P_{[1,1]}P_{[1]} + 2P_{[1]}^3
\approx \\
&= \Tr A_3 + 3\Big(\Tr A_2\Tr A_1 - \Tr(A_2A_1)\Big)
+ \Big((\Tr A_1)^3-3\,\Tr A_1^2\,\Tr A_1 +2\,\Tr A_1^3\Big)\\
&\quad -3 \Big(\Tr A_2  +(\Tr A_1)^2  -\Tr A_1^2\Big)\,\Tr A_1
+ 2(\Tr A_1)^3 \\
&=\Tr\Big(A_3  - 3A_2A_1 + 2A_1^3  \Big).
\end{aligned} 
\end{equation}
%
%

In general the relation
\be
\sum_{k=1}^\infty \frac{\langle \pi^k\rangle_c}{k!} =
\sum_{m=1}^\infty \frac{(-1)^{m+1}}{m} \left(\sum_{k=1}^\infty \frac{\langle \pi^k \rangle}{k!}\right)^m
= \sum_{s=1}^\infty \tilde S_{[1^s]}\{p_k=\langle \pi^k\rangle\}
\ee
should be supplemented by the substitution $\langle\pi^k\rangle = \langle \pi_{s_1}\ldots \pi_{s_k} \rangle$,
which implies symmetrization over all $p_{s_i}$ at the r.h.s.
For example, from $\tilde S_{[1,1,1]} = \frac{1}{6}\Big(p_3-3p_2p_1+2p_1^3\Big)$
\be
\langle\pi_{s_1}\pi_{s_2}\pi_{s_3}\rangle_c =  \langle \pi_{s_1}\pi_{s_2}\pi_{s_3}\rangle
-  \langle\pi_{s_1}\pi_{s_2} \rangle \langle\pi_{s_3}\rangle
-  \langle\pi_{s_2}\pi_{s_3} \rangle \langle\pi_{s_1}\rangle
-  \langle\pi_{s_1}\pi_{s_3} \rangle \langle\pi_{s_2}\rangle
+ 2\langle \pi_{s_1}\rangle\langle \pi_{s_2}\rangle \langle \pi_{s_3}\rangle.
\label{pi3conn}
\ee
Now we should convert $\bra \pi_s\ket$ in a combination of $\sigma_R$ with $|R|=s$
then express $\sigma_R$ through $P_Q$ of the same size $|Q|=s$,
and finally express $P$'s through traces of $A$.
The result of these three operations actually appears implied directly by our initial formula --
all multiple traces disappear and each average gets substituted by a single 
trace of $A$'s:
\begin{equation}
\begin{aligned}
 \langle\pi_{s_1}\pi_{s_2}\pi_{s_3}\rangle_c =
\Tr \Bigl[&A(s_1+s_2+s_3) - A(s_1+s_2)A(s_3) - A(s_2+s_3)A(s_1)
- A(s_3+s_1)A(s_2)\\
&  + A(s_1)A(s_2)A(s_3)+A(s_1)A(s_3)A(s_2)\Bigr].
\end{aligned} 
\label{eq:3pt}
\end{equation}
The last two terms actually coincide 
due to \eqref{eq:sym-AAA}.
Likewise, from $\tilde S_{[1,1,1,1]}=\frac{1}{24}\Big(p_4-4p_3p_1-3p_2^2+12p_2p_1^2 - 6p_1^4\Big)$ we find
\begin{equation}
\begin{aligned}
 &\langle\pi_{s_1}\pi_{s_2}\pi_{s_3}\pi_{s_4}\rangle_c 
=\Tr \Bigl[A(s_1+s_2+s_3+s_4) - A(s_1+s_2+s_3)A(s_4)\\
& - A(s_2+s_3+s_4)A(s_1)
- A(s_3+s_4+s_1)A(s_2)-A(s_4+s_1+s_2)A(s_3)\\
&- A(s_1+s_2)A(s_3+s_4)- A(s_1+s_3)A(s_2+s_4)- A(s_1+s_4)A(s_2+s_3)\\
&+ A(s_1+s_2)(A(s_3)A(s_4) + A(s_4)A(s_3))
+ A(s_1+s_3)(A(s_2)A(s_4)+A(s_4)A(s_2))\\
&+ A(s_1+s_4)(A(s_2)A(s_3)+A(s_3)A(s_2))+ A(s_2+s_3)(A(s_1)A(s_4)+A(s_4)A(s_1))\\
&+ A(s_2+s_4)(A(s_1)A(s_3)+A(s_3)A(s_1))+ A(s_3+s_4)(A(s_1)A(s_2)+A(s_2)A(s_1)) \\
&-A(s_1)A(s_2)A(s_3)A(s_4)
-A(s_1)A(s_2)A(s_4)A(s_3)
-A(s_1)A(s_3)A(s_2)A(s_4)\\
&-A(s_1)A(s_3)A(s_4)A(s_2)
-A(s_1)A(s_4)A(s_2)A(s_3)
-A(s_1)A(s_4)A(s_3)A(s_2)
\Bigr].
\end{aligned} 
\label{pi4conn}
\end{equation}
Since $A$ is a symmetric matrix, some terms in 
the last two lines of \eqref{pi4conn} coincide.
For instance,
\begin{equation}
\begin{aligned}
 \Tr A(s_1)A(s_2)A(s_3)A(s_4)
=\Tr A(s_1)A(s_4)A(s_3)A(s_2).
\end{aligned} 
\end{equation}
\eqref{eq:3pt} and \eqref{pi4conn} agree with the result of 
\cite{Okuyama:2018aij}, as expected.
For illustration, the check of (\ref{pi4conn}) for some set of indices, say $(2,1,1,1)$,
consists of 4  steps:
\begin{itemize}
\item{}
Read   from $\tilde S_4$:
$$\langle \pi_2\pi_1^3 \rangle_c = \langle \pi_2\pi_1^3 \rangle
- 3\langle \pi_2\pi_1^2\rangle\langle \pi_1\rangle - \langle \pi_2 \rangle \langle \pi_1^3 \rangle
- 3\langle \pi_2\pi_1 \rangle \langle \pi_1^2 \rangle
+ 6\langle \pi_2\pi_1 \rangle\langle\pi_1\rangle^2
+ 6 \langle \pi_2 \rangle \langle \pi_1^2 \rangle \langle \pi_1 \rangle
- 6\langle\pi_2 \rangle \langle \pi_1\rangle ^3
$$
\item{}
Express it through averages $\sigma_R = \langle S_R \rangle$ of Schurs:
{\footnotesize
\be
\langle \pi_2\pi_1^3 \rangle_c =
\Big(\sigma_{[5]} + 2\sigma_{[1,4]}+\sigma_{[2,3]}-\sigma_{[1,2,2]}-2\sigma_{[1,1,1,2]}-\sigma_{[1,1,1,1,1]}\Big)
- 3\Big(\sigma_{[4]}+\sigma_{[1,3]}-\sigma_{[1,1,2]}-\sigma_{[1,1,1,1]}\Big)\sigma_{[1]}
-\nn \\
- \Big(\sigma_{[3]}+2\sigma_{[1,2]}+\sigma_{[1,1,1]}\Big)\Big(\sigma_{[2]}-\sigma_{[1,1]}\Big)
- 3\Big(\sigma_{[3]}-\sigma_{[1,1,1]}\Big)\Big(\sigma_{[2]}+\sigma_{[1,1]}\Big)
+ \nn \\
+ 6\Big(\sigma_{[3]}-\sigma_{[1,1,1]}\Big)\sigma_{[1]}^2
+ 6\Big(\sigma_{[2]}-\sigma_{[1,1]}\Big)\Big(\sigma_{[2]}+\sigma_{[1,1]}\Big)\sigma_{[1]}
- 6\Big(\sigma_{[2]}-\sigma_{[1,1]}\Big) \sigma_{[1]}^3 \ \ \ \ \ \
\nn
\ee
}
\item{} Express $\sigma_R$ through $P_Q$ with the help of the Kostka matrix, see (30) in \cite{Morozov:2026jyc}
(we absorb $e^{\# \lambda^2}$ factors into $P$, they will drop out from the final answer,
but we substitute equality by $\approx$, to emphasize that the correlator is not polynomial):
\begin{equation}
\begin{aligned}
 \langle \pi_2\pi_1^3 \rangle_c &\approx
P_{[5]}+3P_{[1,4]}+4P_{[2,3]}+6P_{[1,1,3]}+6P_{[1,2,2]}+6P_{[1,1,1,2]}
\\
&- 3\Big(P_{[4]}+2P_{[1,3]}+2P_{[2,2]}+2P_{[1,1,2]}\Big)P_{[1]}
- 4P_{[3]}P_{[2]} -6P_{[3]}P_{[1,1]}- 6P_{[1,2]}P_{[2]}-6P_{[1,2]}P_{[1,1]}
\\
&+6P_{[3]}P_{[1]}^2+6P_{[1,2]}P_{[1]}^2+ 6P_{[2]}^2P_{[1]}+12P_{[2]}P_{[1,1]}P_{[1]}
-6P_{[2]}P_{[1]}^3-6P_{[2]}P_{[1,1,1]}
\end{aligned} 
\end{equation}
\item{} Substitute expressions for $P$ through traces of $A$ from (41)-(44) of \cite{Morozov:2026jyc}
and obtain the single-trace expression (\ref{pi4conn}):
\be
\langle \pi_2\pi_1^3 \rangle_c =
\Tr \Big(
A_5-3A_4A_1-4A_3A_2+6A_3A_1^2+6A_2^2A_1-
6A_2 A_1^3
\Big).
\ee
\end{itemize}

For the simplest version of (\ref{pi3conn}) these 4 steps are much simpler:\\

\ \ \ $\downarrow$  \ \ \ $\langle \pi^r \rangle_c \sim \tilde S_{[1^r]}\{\langle \pi^k\rangle \}$

\begin{itemize}
\item{} 
$\langle \pi_1^3 \rangle_c := \langle \pi_1^3\rangle - 3\langle \pi_1^2\rangle \langle \pi_1\rangle
+ 2\langle \pi_1\rangle^3$

 $\downarrow$  \ \ \ $U:\ \ \langle\pi_{_Q}\rangle = \sum_{R\vdash Q} \chi_{_{R,Q}}\sigma_{\!_R}$ 

\item{}   $\langle \pi_1^3 \rangle_c  = \sigma_{[3]}+2\sigma_{[1,2]}+\sigma_{[1,1,1]}
-3(\sigma_{[2]}+\sigma_{[1,1]})\cdot \sigma_{[1]} + 2\sigma_{[1]}^3$

$\downarrow$  \ \ \     $V: \ \ \sigma_{_R} \approx \sum_{Q\leq R} K_{R,Q} P_Q$

\item{}  $\langle \pi_1^3 \rangle_c  \approx
P_{[3]}+3P_{[1,2]}+6P_{[1,1,1]} -3(P_{[2]}+2P_{[1,1]})P_{[1]} + 2P_{[1]}^3$

$\downarrow$ \ \ \ $\displaystyle W: \ \  P_R \approx {\rm proj}_R
\left\{ \det\Bigl(1+\sum_{k=1}^\infty t^kA_k\Bigr)\right\}$

\item{} $\langle \pi_1^3 \rangle_c  = \Tr\Big(A_3-3A_2A_1+2A_1^3\Big)$.
\end{itemize}

\noindent
The fact that the last line reproduces the first one, $W\circ V\circ U={\rm Id}$,  
and thus is entirely defined by the polynomial $\tilde S$,
is equivalent to decomposition of Kostka matrix $K_{R,Q}$  in $V$  from (29) of \cite{Morozov:2026jyc}
into the composition of the Schur expansion into time-variables
$U^{-1}$ ($\chi_{_{R,Q}}$ are symmetric-group characters) 
and inverse of determinant expansion $W$, see (6.2) of \cite{Okuyama:2018aij} 
and (40) of \cite{Morozov:2026jyc}.

\section{Comparison to \cite{Morozov:2009uy}}\label{sec:comparison}

Despite connected correlators  satisfy nice single-trace formulas,
their physical significance remains obscure.
Usually connected correlators are introduced to eliminate
pair singularities and concentrate on the less trivial structures of the amplitudes.
Actually there are no singularities in matrix models, but the terminology
and the definitions come from the higher dimensional QFT, where propagators are singular.
For this purpose one should consider subtractions from $\langle \prod_i\Tr X^{k_i}\rangle$,
i.e. first expand our $\langle \prod_i \Tr e^{s_i X}\rangle$
into infinite series, then expand -- and subtract only after that.
We denote this two-step prescription by the double angular brackets in (\ref{expandconn}) below.
The results of this kind were derived in \cite{Morozov:2009uy} and they have their own beauty --
still different from the one, which we emphasize in the present paper.

Additional trick in \cite{Morozov:2009uy} was to use
some peculiar generating functions for correlators, summed over matrix sizes $N$.
The simplest of them is
\be
e_1(z,s) := \sum_{N=1}^\infty z^N\langle \Tr e^{s X}\rangle
= \frac{z}{(1-z)^2}\exp\left( \frac{1+z}{1-z}\cdot \frac{s^2}{2}\right).
\label{e1}
\ee
This can be compared with the fixed $N$ result of
$\bra\Tr e^{s X}\ket$ in \eqref{eq:DG}.
Indeed \eqref{eq:DG} is reproduced from \eqref{e1} 
by using the generating function of the Laguerre polynomials 
\eqref{eq:gen-L}
\be
\frac{z}{(1-z)^2} \exp\left(\frac{zs^2}{1-z}\right)
= \sum_{N=1}^\infty z^N L^1_{N-1}(-s^2).
\label{e1vsL}
\ee

A more important/general claim of \cite{Morozov:2009uy} is that
-- as a consequence of integrability  of the underlying matrix model \cite{Gerasimov:1990is,Morozov:1994hh,Morozov:1995pb,Morozov:2005mz,Morozov:2022bnz,Mironov:1994sf} --
the generating functions
for connected  correlation functions
\be
e_m(z;s_1,s_2,\ldots,s_m) = \sum_{N=1}^\infty z^N
\langle\langle \Tr\,e^{s_1 X}\ldots \Tr\,e^{s_m X}\rangle\rangle_c
\label{expandconn}
\ee
satisfy (5.20) of the JHEP version in \cite{Morozov:2009uy} 
(or (77) of the arXiv version in \cite{Morozov:2009uy}),
\be
z\frac{\p}{\p z}\left(\frac{(1-z)^2}{z}\, e_m(z;s_1,\ldots,s_m) - g_m(z;s_1,\ldots,s_m)\right)
= (s_1+\cdots+ s_m)^2 e_m(z;s_1,\ldots,s_m),
\label{eqforem}
\ee
i.e. that there is an explicit recurrence in $N$ with some free-term function $g$.
Sometime this equation can be solved, like in (\ref{e1}) for $m=1$ when $g_1=0$,
or at $m=2$ -- see (85) in \cite{Morozov:2009uy}, but already this formula is somewhat sophisticated/overloaded.



Actually for ${\bf m=1}$ eq.(\ref{eqforem}) is satisfied
by our
\be
e_1(z;s)=\sum_{N=1}^\infty z^N \Tr_{N\times N} A(s),
\ee
where no double subtractions are made:
\be
z\frac{\p}{\p z}\left(\frac{(1-z)^2}{z} e_1(z;s)  \right) \ \stackrel{(\ref{e1})}{=} \
 z\frac{\p}{\p z}\exp\left( \frac{1+z}{1-z}\cdot \frac{s^2}{2}\right)
= \frac{zs^2}{(1-z)^2}\exp\left( \frac{1+z}{1-z}\cdot \frac{s^2}{2}\right)
= s^2  e_1(z;s).
\ee
Note that in this case normalization does not matter (or is not defined):
$e^{-\frac{s^2}{2}}e_1(z;s)$ satisfies the same equation.
Consistency with Laguerre was already checked in (\ref{e1vsL}),
in terms of equations it means that
\be
\sum_{N=1}^\infty \Big((N-1)z^{N-1}-2Nz^N+(N+1)z^{N+1}\Big)L_{N-1}^1(-s^2)  =
s^2\sum_{N=1}^\infty  z^N L_{N-1}^1(-s^2),
\ee
or
\be
(s^2+2N)L_{N-1}^1(-s^2) = NL_N^1(-s^2)+NL_{N-2}^1(-s^2),
\ee
which is the three-term relation for orthogonal Laguerre polynomials,
see 8.971 in \cite{gradshteyn2014table}.

For ${\bf m\geq 2}$ integrability implications should still be worked out for our connected correlators
(without double brackets).
The physically-artificial nature of our subtraction procedure is obvious already for $m=2$:
in  $\langle \Tr e^{s_1 X} \Tr e^{s_2 X}\rangle_c =
\langle \Tr e^{s_1 X} \Tr e^{s_2 X}\rangle - \langle \Tr e^{s_1 X} 
\rangle\langle \Tr e^{s_2 X}\rangle$
the two items at the r.h.s. are proportional to $e^{\hf(s_1+s_2)^2}$ and $e^{\hf s_1^2+\hf s_2^2}$,
which are different, so that such subtraction would not eliminate any singularities -- if they were present.
Still this definition leads to nice single-trace formulas with $A$ --
and it deserves studying, how {\it integrability} is reflected in their properties,
i.e. expanding the consideration of  \cite{Morozov:2009uy} in this direction as well.

\section{Conclusion and outlook}\label{sec:conclusion}


We provided an expression for the would-be-non-Abelian time-variables,
i.e. the traces of products of non-commuting matrices in \cite{Okuyama:2018aij,Morozov:2026jyc},
through a single Laguerre polynomial $L_{N-1}^1$ --
instead of the huge sets of different extended Laguerres $L_i^\al$.

Described in section \ref{TrueTr} are expressions through $L^1_{N-1}$ of all the ``time-variables'',
appearing in generalized Schur polynomials, which enter the formulas for
Gaussian averages of the group characters $\langle {\rm Tr}_R e^{\cal X}\rangle$
in \cite{Morozov:2026jyc}.
From the point of view of integrability theory 
the most interesting new phenomenon in these formulas is that, say,
\[
 \Tr A(s_1)A(s_2)A(s_3)A(s_4)\ne \Tr A(s_1)A(s_2)A(s_4)A(s_3)
\]
what implies a considerable ``non-abelian'' extension of the set of time-variables. 
According to field/string theory consideration, like in \cite{Okuyama:2018aij},
it seems to be a necessary step in theories with the space-time.
It is a striking fact that this generalization can be made already in matrix models --
thus further extending their role of a simple, still representative prototype 
of the entire string theory.
Hopefully, the simplified expressions for these traces, which are described in this paper,
will help to tame the zoo of these new variables -- and functions, which depend on them. 

Of certain interest is the matching of the single-trace formulas for connected correlators 
in section \ref{sec:conn} with Toda-integrability properties, which we began to discuss in section \ref{sec:comparison}.  
Further work in this direction also looks promising.

\section*{Acknowledgements}
The work of AM is supported by the RSF grant  24-12-00178.
KO is supported 
in part by JSPS Grant-in-Aid for Transformative Research Areas (A)
``Extreme Universe'' 21H05187 and JSPS KAKENHI 25K07300.

\appendix
\section{Alternative derivation of \eqref{eq:tr-conv}}\label{app:alt}
In this appendix, we present an alternative derivation of \eqref{eq:tr-conv}.
Instead of using the generating function of
Laguerre polynomials in \eqref{eq:gen-L},
we can use the following relation 
\begin{equation}
\begin{aligned}
 e^{az}(b+z)^j=\sum_{i=0}^\infty b^{j-i}L_i^{j-i}(-ab)z^i,
\end{aligned} 
\label{eq:gen-L2}
\end{equation}
which can be easily obtained from 8.975-2 in \cite{gradshteyn2014table}.
Then the matrix $A(s)$ in \eqref{eq:Amat-with} is written as
\begin{equation}
\begin{aligned}
 A(s)_{ij}
&=\rt{\frac{i!}{j!}}\oint_{z=0}\frac{dz}{2\pi\ri z^{i+1}}(s+z)^je^{\hf s^2+sz}.
\end{aligned} 
\label{eq:Amat-oint}
\end{equation}
Using this expression, the matrix element of the product $A(s_1)A(s_2)$
becomes
\begin{equation}
\begin{aligned}
 \bigl(A(s_1)A(s_2)\bigr)_{ij}&=\sum_{k=0}^{N-1}A(s_1)_{ik}A(s_2)_{kj}\\
&=\rt{\frac{i!}{j!}}\sum_{k=0}^{N-1}
\oint_{z_1=0}\frac{dz_1}{2\pi\ri z_1^{i+1}}(s_1+z_1)^k 
\oint_{z_2=0}\frac{dz_2}{2\pi\ri z_2^{k+1}}(s_2+z_2)^j \prod_{l=1,2}e^{\hf s_l^2+s_lz_l}\\
&=\rt{\frac{i!}{j!}}
\oint_{z_1=0}\frac{dz_1}{2\pi\ri z_1^{i+1}}
\oint_{z_2=0}\frac{dz_2}{2\pi\ri }\left[\left(\frac{s_1+z_1}{z_2}\right)^N-\underline{1}
\right]
\frac{(s_2+z_2)^j}{s_1+z_1-z_2}\prod_{l=1,2}e^{\hf s_l^2+s_lz_l}.
\end{aligned} 
\end{equation}
By the same argument as in section \ref{sec:A1}, the 
underlined $\underline{1}$ does not contribute to the integral 
and hence it can be omitted.
Repeating this computation,
we find
\begin{equation}
\begin{aligned}
 \Tr A(s_1)A(s_2)\cdots A(s_m)
=\prod_{l=1}^m \oint_{z_l=0}\frac{dz_l}{2\pi\ri}e^{\hf s_l^2+s_lz_l}
\left(\frac{s_l+z_l}{z_l}\right)^N\frac{1}{s_l+z_l-z_{l+1}},
\end{aligned} 
\label{eq:AAA-oint}
\end{equation}
where $z_{m+1}=z_1$.
Note that this expression for $m=2$ 
has already appeared in \cite{brezin-hikami}.
The product of the last factor of \eqref{eq:AAA-oint} is 
expanded as
\begin{equation}
\begin{aligned}
 \prod_{l=1}^m\frac{1}{s_l+z_l-z_{l+1}}&=\frac{1}{s_1+\cdots +s_m}
\sum_{k=1}^m\prod_{\substack{j=1\\ j\ne k}}^{m}\frac{1}{s_j+z_j-z_{j+1}}.
\end{aligned} 
\label{eq:prop}
\end{equation}
For instance, the $m=2$ case reads
\begin{equation}
\begin{aligned}
 \frac{1}{(s_1+z_1-z_2)(s_2+z_2-z_1)}=\frac{1}{s_1+s_2}\left(
\frac{1}{s_1+z_1-z_2}+\frac{1}{s_2+z_2-z_1}\right).
\end{aligned} 
\end{equation}
For general $m$, one can prove \eqref{eq:prop} 
by multiplying the following decomposition of $1$ to the left-hand side
of \eqref{eq:prop}
\begin{equation}
\begin{aligned}
 1=\frac{1}{s_1+\cdots +s_m}\sum_{k=1}^m (s_k+z_k-z_{k+1}).
\end{aligned} 
\label{eq:ptep-id2}
\end{equation}
We can exponentiate $(s_j+z_j-z_{j+1})^{-1}$ in 
\eqref{eq:prop} via the 
Schwinger representation
\begin{equation}
\begin{aligned}
 \frac{1}{s_j+z_j-z_{j+1}}=\int_0^\infty dx_j 
e^{-(s_j+z_j-z_{j+1})x_j}.
\end{aligned} 
\end{equation}
This holds for $s_j>0$, since $z_j$ can be taken to be sufficiently small
$|z_j|\ll1$.
Then \eqref{eq:prop} becomes
\begin{equation}
\begin{aligned}
 \prod_{l=1}^m\frac{1}{s_l+z_l-z_{l+1}}&=\frac{1}{\sum_{k=1}^m s_k}
\prod_{j=1}^{m-1}\int_0^\infty dx_j e^{-(s_j+z_j-z_{j+1})x_j}+(\text{cyclic permutation}).
\end{aligned} 
\label{eq:prop-Sch}
\end{equation}
Finally, plugging \eqref{eq:prop-Sch} into \eqref{eq:AAA-oint}
and using \eqref{eq:gen-L2} again, we
arrive at our desired relation \eqref{eq:tr-conv}.

\bibliography{cite}
\bibliographystyle{utphys}

\end{document}